\documentclass[preprint]{revtex4}

\usepackage{amsfonts}
\usepackage{amssymb}
\usepackage{latexsym}
\usepackage{graphicx}
\usepackage{epsfig,subfig}
\usepackage{epstopdf}
\usepackage[active]{srcltx}

\begin{document}

\title{Jaynes-Cummings model in a finite Kerr medium }
\author{A. Mart\'{\i}n Ruiz}
\affiliation{Instituto de Ciencias Nucleares, Universidad Nacional Aut\'{o}noma de M\'{e}%
xico, 04510 M\'{e}xico, D.F., Mexico}
\author{Alejandro Frank}
\affiliation{Instituto de Ciencias Nucleares, Universidad Nacional Aut\'{o}noma de M\'{e}%
xico, 04510 M\'{e}xico, D.F., Mexico}
\affiliation{Centro de Ciencias de la Complejidad, Universidad Nacional Aut\'{o}noma de M%
\'{e}xico, 04510 M\'{e}xico, D.F., Mexico}
\author{L. F. Urrutia}
\affiliation{Instituto de Ciencias Nucleares, Universidad Nacional Aut\'{o}noma de M\'{e}%
xico, 04510 M\'{e}xico, D.F., Mexico}
\affiliation{Facultad de F\'{\i}sica, Pontificia Universidad Cat\'{o}lica de Chile,
Casilla 306, Santiago 22, Chile}

\begin{abstract}
We introduce a spin model which exhibits the main properties of a Kerr
medium to describe an intensity dependent coupling between a two-level atom
and the radiation field. We select a unitary irreducible representation of
the $su(2)$ Lie algebra such that the number of excitations of the field is
bounded from above. We analyze the behavior of both the atomic and the field
quantum properties and its dependence on the maximal number of excitations.
\end{abstract}

\maketitle

\section{Introduction}

The Jaynes-Cummings model (JCM) \cite{1} is one of the simplest quantum
systems describing the interaction of matter with electromagnetic radiation
employing a Hamiltonian of a two-level atom coupled to a single bosonic
mode. Due to its importance in laser physics and quantum optics, this model
has been the subject of many recent investigations, both theoretical and
experimental \cite{2}. It has been observed \cite{3} that the temporal
behavior of this system is very sensitive to the statistical properties of
the radiation field in the initial state, revealing pure quantum features
which have no classical counterpart, such as vacuum Rabi oscillations \cite%
{3A}, collapse and revival phenomena \cite{4} and squeezing of the radiation
field \cite{4A}. These predictions have been verified experimentally \cite{5}%
.

The JCM has been generalized in different ways. In some cases, the
interaction between the atom and the radiation field is no longer linear in
the field variables, i.e. intensity dependent coupling has been considered 
\cite{6}. Other investigations relate the JC Hamiltonian to $u\left(
1|1\right) $ and $osp\left( 2|2\right) $ superalgebras \cite{7}. Most of the
nonlinear generalizations of the JCM are made by using appropriate $q$%
-deformed oscillators \cite{8,9}. In this framework, the application of the
corresponding quantum algebras has proved useful in obtaining exactly
soluble models \cite{10,11}. Additional generalizations of the JCM can be
found in \cite{11B}.

Nonlinear phenomena are ubiquitous in quantum optics. One of the simplest
phenomenon is the Kerr effect, which occurs when the refractive index of a
medium varies with the number of excitations of the field. Quantum
descriptions of optical fields propagating in a Kerr medium reveal a number
of interesting features, such as photon antibunching, squeezing and the
formation of Schr\"{o}dinger cats \cite{12,14}, which have no classical
analogues. The Kerr medium has been recently considered in the framework of $%
q$-deformed oscillators \cite{15} and the Moyal phase space representation 
\cite{16}. Some authors have considered a system where a two-level atom is
surrounded by a Kerr medium using a special case of $q$-deformed oscillators 
\cite{17}.

In this paper, we use an $SU(2)$ spin model, which exhibits the main
properties of the Kerr medium \cite{18}, to describe an intensity dependent
coupling between a two-level atom and the radiation field in the framework
of the JCM. The main feature of the model is that it is formulated on a
finite dimensional Hilbert space, so that the number of excitations of the
field is bounded from above. In the limit when the dimension of the Hilbert
space becomes infinite, the $su\left( 2\right) $ algebra of the spin
operators contracts to the Heisenberg-Weyl algebra of boson operators and
the model coincides with the usual JCM.

\section{A Kerr Hamiltonian arising from an $SU(2)$ spin system}

It is well known that the Heisenberg-Weyl algebra can be obtained by a
contraction of the $su\left( 2\right) $ algebra \cite{18A}. Taking this
limit as a motivation, let us start by considering the $su\left( 2\right) $
commutation relations 
\begin{equation}
\lbrack \hat{S}_{+},\hat{S}_{-}]=2\hat{S}_{3},~\ [\hat{S}_{3},\hat{S}_{\pm
}]=\pm \hat{S}_{\pm },~\ \hat{S}_{\pm }=\hat{S}_{1}\pm i\hat{S}_{2}.
\end{equation}%
In a given\textbf{\ }unitary irreducible representation $j$, we introduce
the operators%
\begin{equation}
\hat{b}=\frac{\hat{S}_{-}}{\sqrt{2j}},~\ ~\hat{b}^{\dag }=\frac{\hat{S}_{+}}{%
\sqrt{2j}},~~\ \hat{n}=\hat{S}_{3}+j,  \label{DEFOP}
\end{equation}%
which act on a $\left( 2j+1\right) $-dimensional Hilbert space of a particle
with spin $j$ \cite{18}. The $su\left( 2\right) $ algebra in this
representation is thus written as%
\begin{equation}
\lbrack \hat{b},\hat{n}]=\hat{b},~~\ [\hat{b}^{\dag },\hat{n}]=-\hat{b}%
^{\dag },~~\ [\hat{b},\hat{b}^{\dag }]=1-\frac{\hat{n}}{j},  \label{ALG}
\end{equation}%
which allows the interpretation of $b^{\dagger }$ ($b$) as creation
(annihilation) operators for the quanta labeled by the number operator $\hat{%
n}$\textbf{. } The oscillator-like\textbf{\ }Hamiltonian, in terms of the
spin\textbf{\ }$j$ operators, is\textbf{\ } 
\begin{equation}
\hat{H}=\frac{\hbar \omega }{4j}(\hat{S}_{+}\hat{S}_{-}+\hat{S}_{-}\hat{S}%
_{+})=\hbar \omega \left( \hat{n}+\frac{1}{2}-\frac{\hat{n}^{2}}{2j}\right) .
\end{equation}%
The analogy of this expression with the usual optical Kerr Hamiltonian is
evident, but the main difference is that the spin number excitation is
bounded from above: $n\leq 2j$. This is a consequence of the angular
momentum condition $-j\leq m\leq +j$, once it is written in terms of the
eigenvalues of the number operator $\hat{n}$. Clearly the energy spectrum
has the twofold degeneracy $E\left( n\right) =E\left( 2j-n\right) $. Note
that although this Hamiltonian can be easily obtained in the framework of $q$%
-deformed oscillators, we use an algebraic model which possesses a physical
interpretation of quantum deformation based on the fundamental $SU\left(
2\right) $ group. The standard harmonic oscillator limit is obtained when $%
j\rightarrow \infty $, in which case $\hat{b}$ and $\hat{b}^{\dag }$ turn
out to be\textbf{\ } the usual bosonic operators acting on an infinite
dimensional Hilbert space, i.e., the nonlinearity and the maximum number of
excitations disappear. We observe that the relation between the number
operator and the corresponding creation-annihilation operators is\textbf{\ } 
\begin{equation}
\hat{n}=\frac{(2j+1)}{2}\left[ 1-\sqrt{1-\frac{8j}{(2j+1)^{2}}\hat{b}^{\dag
}b}\right] ,
\end{equation}%
which leads to the usual relation $\hat{n}=\hat{b}^{\dag }b$ when $%
j\rightarrow \infty $.

The solution of Heisenberg's equation of motion for spin operators is 
\begin{eqnarray}
\hat{S}_{1}(t) &=&e^{i\frac{\omega t}{2j}}[\cos (\hat{\Omega}t)\hat{S}%
_{1}-\sin (\hat{\Omega}t)\hat{S}_{2}],  \nonumber \\
\hat{S}_{2}(t) &=&e^{i\frac{\omega t}{2j}}[\sin (\hat{\Omega}t)\hat{S}%
_{1}+\cos (\hat{\Omega}t)\hat{S}_{2}],  \nonumber \\
\hat{S}_{3}(t) &=&\hat{S}_{3},
\end{eqnarray}%
where $\hat{\Omega}=\omega (1-\frac{\hat{n}}{j})$. Thus, the time evolution
of $\vec{S}\left( t\right) $ is a rotation around the $z$ axis, but the
precession frequency depends on the excitation number operator. This result
is in complete analogy with that obtained in the case of the Kerr medium 
\cite{19,20}.

In molecular physics, this model correspond to a diatomic molecule
approximated by a Morse potential and the eigenstates correspond to the $%
U(2)\supset SU(2)$ symmetry-adapted basis \cite{21,22}. It was shown that
this type of Hamiltonian describes the interaction of a collective atomic
system with the off-resonant radiation field in a dispersive cavity \cite{23}%
.

\section{SU(2) Coherent states}

Nonlinear coherent states (NCSs) have been discussed in different approaches
and arriving to different states \cite{24,25}. The most representative\ ones
are constructed\textbf{\ } in the framework of $q$-deformed oscillators \cite%
{26}, in which NCSs are eigenstates of the $q$-annihilation operator. A
generalization of the $SU(2)$ coherent states given by a multiphoton
Holstein-Primakoff transformation is discussed in Ref.\cite{26A}, and their
physical consequences are studied in the framework of the standard JC model.
Here we describe the $SU(2)$ coherent states corresponding to the single
photon case in the SU(2) spin model and use them to study the modified JC
model to be defined in the next section.

The group-theoretical coherent state (see ref.\cite{23A}) for the unitary
irreducible representation of the $su(2)$ Lie algebra corresponding to Eq.(%
\ref{DEFOP}) is:%
\begin{equation}
\left\vert \alpha \right\rangle =e^{\alpha \hat{S}_{+}-\alpha ^{\ast }\hat{S}%
_{-}}\left\vert j,-j\right\rangle ,  \label{DEFCOHSU2}
\end{equation}%
where $\alpha =\frac{\theta }{2}e^{-i\phi }$, $0\leq \theta \leq \pi $, $%
0\leq \phi \leq 2\pi $ and the ladder operators $\hat{S}_{\pm }$ select the
vacuum state $\left\vert 0\right\rangle $ from the states $\left\vert
j,m\right\rangle $ as usual: $\hat{S}_{-}\left\vert 0\right\rangle =\hat{S}%
_{-}\left\vert j,-j\right\rangle =0$. The natural phase space is the sphere
of radius $j$; the spin coherent states are thus represented by spots on the
sphere. This dynamics leads to Schr\"{o}dinger cat states on the sphere \cite%
{18}, i.e., a superposition of several spots located \textquotedblleft
far\textquotedblright\ from each other. In the harmonic limit, the sphere
opens to the phase plane and the model coincides with the quantum harmonic
oscillator or, through a renormalization, with the usual Kerr medium.

Using the disentangling theorem for $SU(2)$ operators, we can rewrite Eq.(%
\ref{DEFCOHSU2}) in the following form%
\begin{equation}
\left\vert \xi \right\rangle =(1+\left\vert \xi \right\vert
^{2})^{-j}\sum_{m=-j}^{+j}\left( 
\begin{array}{c}
2j \\ 
j+m%
\end{array}%
\right) ^{\frac{1}{2}}\xi ^{\left( j+m\right) }\left\vert j,m\right\rangle
,\;\;\;
\end{equation}%
where $\xi =\tan \left( \frac{\theta }{2}\right) e^{-i\phi }$. Then, the
photon number distribution of the field prepared in the state $\left\vert
\xi \right\rangle $ is%
\begin{equation}
\mathcal{P}_{n}\left( \left\vert \xi \right\vert ^{2}\right) =\left( 
\begin{array}{c}
2j \\ 
n%
\end{array}%
\right) \frac{\left\vert \xi \right\vert ^{2}}{\left( 1+\left\vert \xi
\right\vert ^{2}\right) ^{2j}},  \label{FOTDIST}
\end{equation}%
and the mean photon number is%
\begin{equation}
\left\langle \hat{n}\right\rangle =2j\frac{\left\vert \xi \right\vert ^{2}}{%
1+\left\vert \xi \right\vert ^{2}}.
\end{equation}%
This result shows that $\left\langle \hat{n}\right\rangle $ is bounded from
above. It is convenient to rewrite Eq. (\ref{FOTDIST}) in the following form%
\begin{equation}
\mathcal{P}_{n}\left( \chi \right) =\left( 
\begin{array}{c}
2j \\ 
n%
\end{array}%
\right) \chi ^{n}\left( 1-\chi \right) ^{2j-n},
\end{equation}%
which is a binomial distribution in $n$ with parameter $\chi \equiv \frac{%
\left\langle \hat{n}\right\rangle }{2j}\in \left[ 0,1\right] $. Evidently $%
\mathcal{P}_{n}\left( \chi \right) $ converges towards the Poisson
distribution if $\left\langle \hat{n}\right\rangle $ remains fixed when $%
2j\rightarrow \infty $. One of the most interesting features of the $SU(2)$
coherent states is that they exhibit squeezing which depends on $2j$.

\section{Two-level atom surrounded by a Kerr medium.}

We start with a brief review of the standard formulation of the JCM. In the
dipole and rotating wave approximation, the JC Hamiltonian for a system of a
single atom interacting with a single mode is given by \cite{1}%
\begin{equation}
\mathcal{H}_{JC}=\frac{1}{2}\hbar \omega (\hat{a}^{\dag }\hat{a}+\hat{a}\hat{%
a}^{\dag })+\frac{1}{2}\hbar \omega _{0}\hat{\sigma}_{3}+\hbar \lambda (\hat{%
\sigma}_{+}\hat{a}+\hat{a}^{\dag }\hat{\sigma}_{-}),
\end{equation}%
where $\frac{1}{2}\hbar \omega (\hat{a}^{\dag }\hat{a}+\hat{a}\hat{a}^{\dag
})$ and $\frac{1}{2}\hbar \omega _{0}\hat{\sigma}_{3}$ are the well-known
energy operators for the field and atom, respectively. Here $\omega $ is the
field mode frequency and $\omega _{0}$ is the atomic transition frequency.
The coupling between the atom and the radiation field is described by $\hbar
\lambda (\hat{\sigma}_{+}\hat{a}+\hat{a}^{\dag }\hat{\sigma}_{-})$, where $%
\lambda $ is a coupling constant. Besides $\hat{a}$ and $\hat{a}^{\dag }$
are the usual bosonic creation and annihilation operators for photons in the
mode, which obey the Heisenberg-Weyl algebra. $\hat{\sigma}_{+}$ and $\hat{%
\sigma}_{-}$ are the usual rising and lowering operators describing the
fermionic two-level atom and $\hat{\sigma}_{3}$ is the atomic inversion
operator, which follow the standard pseudo-spin algebra.

Now, let us introduce the spin model discussed in Sec. II into the JC
Hamiltonian to describe a two-level atom surrounded by a Kerr medium. This
is achieved by replacing the usual bosonic creation and annihilation
operators by the correspondent spin-$j$ operators, i.e. we now consider%
\textbf{\ }%
\begin{equation}
\mathcal{H}_{JC}=\frac{1}{4j}\hbar \omega (\hat{S}_{+}\hat{S}_{-}+\hat{S}_{-}%
\hat{S}_{+})+\frac{1}{2}\hbar \omega _{0}\hat{\sigma}_{3}+\frac{\hbar
\lambda }{\sqrt{2j}}(\hat{\sigma}_{+}\hat{S}_{-}+\hat{S}_{+}\hat{\sigma}%
_{-}).  \label{HAMMOD}
\end{equation}%
Next, we will concentrate on studying the dynamics of the system. We first
split the Hamiltonian into two parts : $\mathcal{H}_{0}$ \textbf{(}the
energy operator in the absence of interaction\textbf{)} and $\hat{\mathcal{V}%
}$ (the coupling interaction), such that 
\begin{equation}
\mathcal{H}_{JC}=\mathcal{H}_{0}+\hat{\mathcal{V}},
\end{equation}%
where 
\begin{eqnarray}
\mathcal{H}_{0} &=&\frac{1}{4j}\hbar \omega (\hat{S}_{+}\hat{S}_{-}+\hat{S}%
_{-}\hat{S}_{+})+\frac{1}{2}\hbar \omega _{0}\hat{\sigma}_{3}, \\
\hat{\mathcal{V}} &\mathcal{=}&\frac{\hbar \lambda }{\sqrt{2j}}(\hat{\sigma}%
_{+}\hat{S}_{-}+\hat{S}_{+}\hat{\sigma}_{-}).
\end{eqnarray}%
In the interaction picture generated by $\mathcal{H}_{0}$, the Hamiltonian
of the system is%
\begin{equation}
\hat{\mathcal{V}}_{\mathcal{I}}\left( t\right) =e^{i\frac{\mathcal{H}_{0}t}{%
\hbar }}\hat{\mathcal{V}}e^{-i\frac{\mathcal{H}_{0}t}{\hbar }}.
\end{equation}%
It is clear that $\hat{\sigma}_{\pm }$ and $\hat{S}_{\pm }$ become
time-dependent by means of the Heisenberg equations of motion. Then the
interaction picture Hamiltonian becomes 
\begin{equation}
\hat{\mathcal{V}}_{\mathcal{I}}\left( t\right) =\frac{\hbar \lambda }{\sqrt{%
2j}}{(\hat{\sigma}}_{+}e^{i\hat{\Omega}_{\hat{n}}t}\hat{S}_{-}+\hat{S}%
_{+}e^{-i\hat{\Omega}_{\hat{n}}t}\hat{\sigma}_{-}),
\end{equation}%
where 
\begin{equation}
\hat{\Omega}_{\hat{n}}=\omega _{0}-\omega \left( 1-\frac{\hat{n}}{j}-\frac{1%
}{2j}\right)
\end{equation}%
\textbf{\ }gives rise to the generalized detuning frequency, which depends
on the excitation number operator.

In order to solve the equation of motion in the interaction picture, we
first observe that $\hat{\mathcal{V}}_{\mathcal{I}}$ describes processes
where a photon in the mode is annihilated while the atom is excited, or vice
versa. Then, at any time $t$, the state of the system $\left\vert \psi
\left( t\right) \right\rangle $ is expanded in terms of the states $%
\left\vert n\right\rangle \otimes \left\vert +\right\rangle $ and $%
\left\vert n+1\right\rangle \otimes \left\vert -\right\rangle $, where $%
\left\vert n\right\rangle $ is an eigenstate of the excitation number
operator and $\left\vert +\right\rangle $ and $\left\vert -\right\rangle $
denote the atom in the excited and ground state, respectively. Thus%
\begin{equation}
\left\vert \psi \left( t\right) \right\rangle =\sum_{n=0}^{2j}\left[
a_{n}\left( t\right) \left\vert n\right\rangle \otimes \left\vert
+\right\rangle +b_{n}\left( t\right) \left\vert n+1\right\rangle \otimes
\left\vert -\right\rangle \right] .
\end{equation}%
The equation of motion for the state of the system in the interaction
picture, i.e. $i\hbar \frac{\partial }{\partial t}\left\vert \psi \left(
t\right) \right\rangle =\hat{\mathcal{V}}_{\mathcal{I}}\left( t\right)
\left\vert \psi \left( t\right) \right\rangle $, gives the following simple
coupled set of differential equations for the probability amplitudes%
\begin{eqnarray}
i\dot{a}_{n} &=&\lambda \sqrt{\left( n+1\right) \left( 1-\frac{n}{2j}\right) 
}e^{i\Omega _{n}t}b_{n}, \\
i\dot{b}_{n} &=&\lambda \sqrt{\left( n+1\right) \left( 1-\frac{n}{2j}\right) 
}e^{-i\Omega _{n}t}a_{n},
\end{eqnarray}%
whose general solution is%
\begin{eqnarray}
&& a_{n}\left( t\right)=\left\{ a_{n}\left( 0\right) \left[ \cos \left( 
\frac{\Gamma _{n}t}{2}\right) -i\frac{\Omega _{n}}{\Gamma _{n}}\sin \left( 
\frac{\Gamma _{n}t}{2}\right) \right] -ib_{n}\left( 0\right) \sqrt{1-\frac{%
\Omega _{n}^{2}}{\Gamma _{n}^{2}}}\sin \left( \frac{\Gamma _{n}t}{2}\right)
\right\} e^{i\Omega _{n}t},  \nonumber \\
&& b_{n}\left( t\right)=\left\{ b_{n}\left( 0\right) \left[ \cos \left( 
\frac{\Gamma _{n}t}{2}\right) +i\frac{\Omega _{n}}{\Gamma _{n}}\sin \left( 
\frac{\Gamma _{n}t}{2}\right) \right] -ia_{n}\left( 0\right) \sqrt{1-\frac{%
\Omega _{n}^{2}}{\Gamma _{n}^{2}}}\sin \left( \frac{\Gamma _{n}t}{2}\right)
\right\} e^{-i\Omega _{n}t},  \nonumber \\
\end{eqnarray}%
where $a_{n}\left( 0\right) $ and $b_{n}\left( 0\right) $ are determined
from the initial conditions of the system and 
\begin{equation}
\Gamma _{n}=\sqrt{\Omega _{n}^{2}+4\lambda ^{2}\left( n+1\right) \left( 1-%
\frac{n}{2j}\right) }
\end{equation}%
is the generalized Rabi frequency.

This set of equations gives us the general solution of the problem. In order
to calculate some physical quantities of interest, we still need to specify
the initial photon number distribution of the field.

Exact solutions of generalized extensions of the JCM in the Schr\"{o}dinger
picture have been calculated in Ref.\cite{11B}. Our modified algebra (Eq.\ref%
{ALG}) is also included in this generalization with the appropriate choice
of the structure functions. Unlike this approach, here we use a unitary
irreducible representation of the $su(2)$ algebra to describe a Kerr medium,
taking advantage of algebraic methods. In addition, we now explore the
physical consequences of the model.

\section{Quantum properties}

Since the JCM is a composite system, which includes the field and the atom
sectors, the quantum properties can be described using the joint density
operator\textbf{\ }%
\begin{equation}
\hat{\rho}_{\mathcal{AF}}\left( t\right) =\left\vert \psi \left( t\right)
\right\rangle \left\langle \psi \left( t\right) \right\vert ,
\end{equation}%
where the subscript $\mathcal{A}$\ and $\mathcal{F}$\ refers to the atom and
the field contributions.

The reduced density operators are then sufficient for calculating the
averages of any dynamical variables that belong exclusively to one of the
components. The reduced density operators are%
\begin{equation}
\hat{\rho}_{\mathcal{A}}\left( t\right) =Tr_{\mathcal{F}}\left[ \hat{\rho}_{%
\mathcal{AF}}\left( t\right) \right] ,~\ \ \ \hat{\rho}_{\mathcal{F}}\left(
t\right) =Tr_{\mathcal{A}}\left[ \hat{\rho}_{\mathcal{AF}}\left( t\right) %
\right] ,
\end{equation}%
with which we are able to evaluate the expectation values of any atomic
operator $\mathcal{O}_{\mathcal{A}}$ (e.g. population inversion operator $%
\hat{\sigma}_{3}$) and of any field operator $\mathcal{O}_{\mathcal{F}}$
(e.g. excitation number operator $\hat{n}$) using the expressions 
\begin{equation}
\left\langle \mathcal{O}_{\mathcal{A}}\right\rangle =Tr_{\mathcal{A}}\left[ 
\hat{\rho}_{\mathcal{A}}\left( t\right) \mathcal{O}_{\mathcal{A}}\right] ,~\
\ \left\langle \mathcal{O}_{\mathcal{F}}\right\rangle =Tr_{\mathcal{F}}\left[
\hat{\rho}_{\mathcal{F}}\left( t\right) \mathcal{O}_{\mathcal{F}}\right] .
\label{MEANVAL}
\end{equation}%
The matrix density of the atom has dimension two and, unlike the usual JCM,
the matrix density of the field has dimension $(2j+1)$.

\subsection{Atomic population inversion}

Many interesting atomic quantum effects have been observed in the context of
the standard JCM. Perhaps the most notable is the periodic transfer of
population between the ground state and the excited state. These transitions
constitute\textbf{\ } the so-called collapse and revival phenomena (CR).
Physically, CR occur because an atom within a cavity undergoes reversible
spontaneous emission, as it repeatedly emits and then reabsorbs radiation 
\cite{27}.

The general expression for the atomic population inversion is%
\begin{equation}
\left\langle \hat{\sigma}_{3}\left( t\right) \right\rangle =Tr_{\mathcal{A}}%
\left[ \hat{\rho}_{\mathcal{A}}\left( t\right) \hat{\sigma}_{3}\right]
=\sum_{n=0}^{2j}\left[ \left\vert a_{n}\left( t\right) \right\vert
^{2}-\left\vert b_{n}\left( t\right) \right\vert ^{2}\right] ,
\end{equation}%
where we have used \ Eq. (\ref{MEANVAL}). We can see that the temporal
evolution of $\left\langle \hat{\sigma}_{3}\left( t\right) \right\rangle $
is essentially a result of a summation of probabilities at different Rabi
frequencies.

In order to compare the differences between the usual treatment and the spin
model, let us consider the atom initially in the excited state and the
initial photon number distribution described by the$\;SU\left( 2\right) $
coherent states, i. e. $\left\vert b_{n}\left( 0\right) \right\vert ^{2}=0$
and $\left\vert a_{n}\left( 0\right) \right\vert ^{2}=\mathcal{P}_{n}\left(
\chi \right) $. Thus%
\begin{equation}
\left\langle \hat{\sigma}_{3}\left( t\right) \right\rangle
=\sum_{n=0}^{2j}\left( 
\begin{array}{c}
2j \\ 
n
\end{array}%
\right) \chi ^{n}\left( 1-\chi \right) ^{2j-n}\left[ \frac{\Omega _{n}^{2}}{%
\Gamma _{n}^{2}}+\left( 1-\frac{\Omega _{n}^{2}}{\Gamma _{n}^{2}}\right)
\cos \left( \Gamma _{n}t\right) \right] .  \label{CR}
\end{equation}%
We observe that this solution converges towards the usual solution of JCM as 
$2j$ approaches infinity, i.e. when the $su\left( 2\right) $ algebra
contracts to the Heisenberg-Weyl algebra. Simultaneously, the photon number
distribution of the field converges towards the Poisson distribution when $%
2j\chi $ remains fixed.

It is well known that the system under consideration is sensitive to the
statistical properties of the electromagnetic field. In the $q$-deformed
extensions of the JCM, it is usually considered that the field is initially
prepared in a $q$-deformed coherent state (an eigenstate of the $q\,$%
-deformed annihilation operator). On the other hand, in Ref.\cite{11B} the
field is prepared in the multipothon Holstein-Primakoff $SU(2)$ coherent
state in the framework of the standard JCM. In this work we consider both
situations: an $SU(2)$ spin model to describe an intensity dependent
coupling, together with its associated coherent state as the initial photon
number distribution.

\begin{figure}
\vspace {0.5 cm}
\begin{center}
\includegraphics[scale=0.8, natwidth=640, natheight=480]{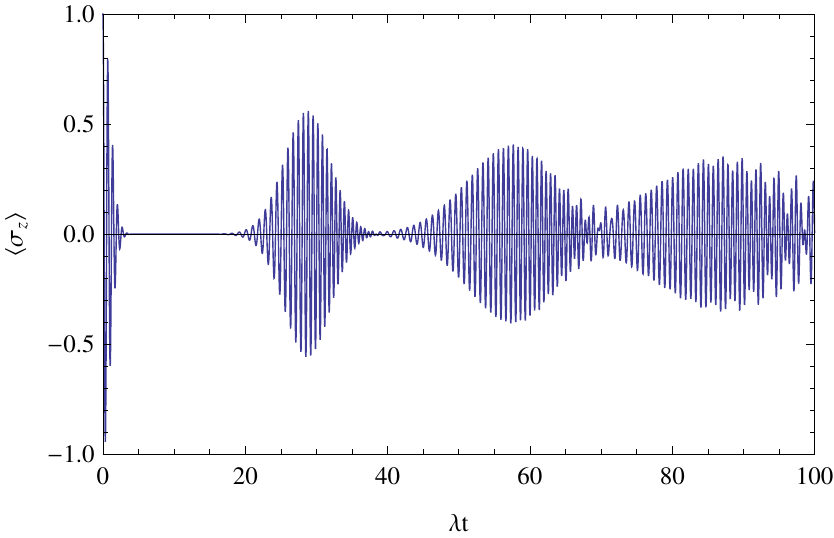}
\caption{Plot of the temporal evolution of the atomic inversion $\left\langle \hat{\sigma}%
_{3}\left( t\right) \right\rangle $ in the standard JCM in the exact resonant case. The atom is initially in the excited state and the field is initially prepared in the standard coherent state with $\left\langle \hat{n}\right\rangle =20$ photons on average.}
\end{center}
\end{figure}

Numerical results of the atomic inversion in the exact resonant case $\left(
\omega =\omega _{0}=\lambda \right) $, with $\left\langle \hat{n}%
\right\rangle =20$ photons on average and$\;2j\rightarrow \infty $, are
shown in Fig. 1. It can be seen that the temporal evolution exhibits
periodic collapses and revivals. This phenomenon also takes place in the
spin model; however, in this case the structure is more complex due to its
dependence on the maximum number of spin excitations $2j$. Figures (2a),
(2b) and (2c) show the temporal evolution of $\left\langle \hat{\sigma}%
_{3}\left( t\right) \right\rangle $ for $\left\langle \hat{n}\right\rangle
=20$ photons on average and $2j=1000,~100$ and$~50$, respectively. We
observe in Fig. 2a that the sequence of CR is essentially the same as in the
limiting case (Fig. 1), but the scaled time $\lambda t_{R}$ needed for the
largest revival of the initial value $\left\langle \hat{\sigma}_{3}\left(
0\right) \right\rangle $ depends on the maximum occupation $2j$. An estimate
of the scaled time ($2j>>1$) in this case gives%
\begin{equation}
\lambda t_{R}\approx \frac{\pi }{\sqrt{\chi ^{2}+\left( 2+2j\chi \right)
\left( 1-\chi \right) }-\sqrt{\chi ^{2}+\left( 1+2j\chi \right) \left(
1-\chi \right) }},
\end{equation}%
which agrees with the numerical results. It can be seen that when $2j$
decreases the structure starts to deteriorate (see Fig. (2b) and (2c)). Due
to the symmetry of the photon number distribution, i.e. $\mathcal{P}%
_{n}\left( \chi \right) =\mathcal{P}_{2j-n}\left( 1-\chi \right) $, periodic
collapses and revivals also take place for any value of $\chi $, except for
the limiting case $\chi =1$. In this case, which does not occur in the
standard JCM, the leading term in the sum of Eq.(\ref{CR}) is that for which 
$n=2j$ and therefore $\left\langle \hat{\sigma}_{3}\left( t\right)
\right\rangle =1$, i.e. the atom remains in the excited state.

\begin{figure}%
\vspace {0.5 cm}
\begin{center}
\subfloat[]{\includegraphics[scale=0.8, natwidth=640, natheight=480]{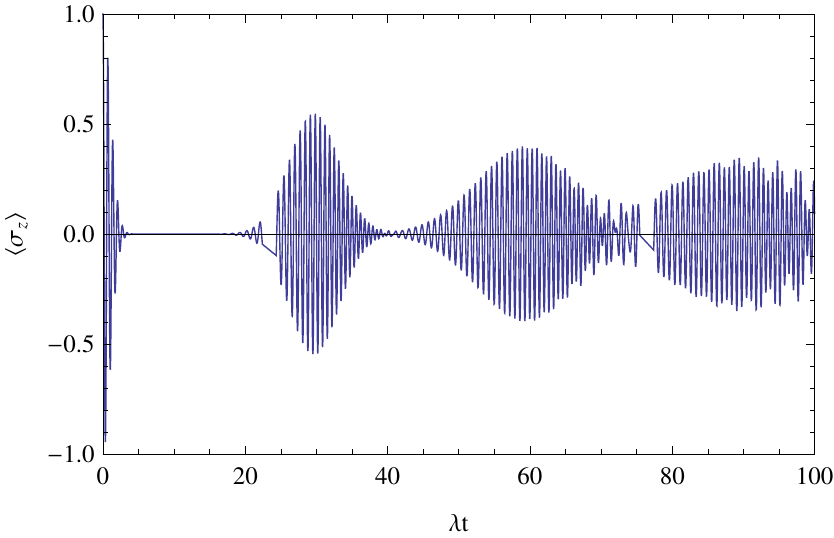}}\qquad
\subfloat[]{\includegraphics[scale=0.8, natwidth=640, natheight=480]{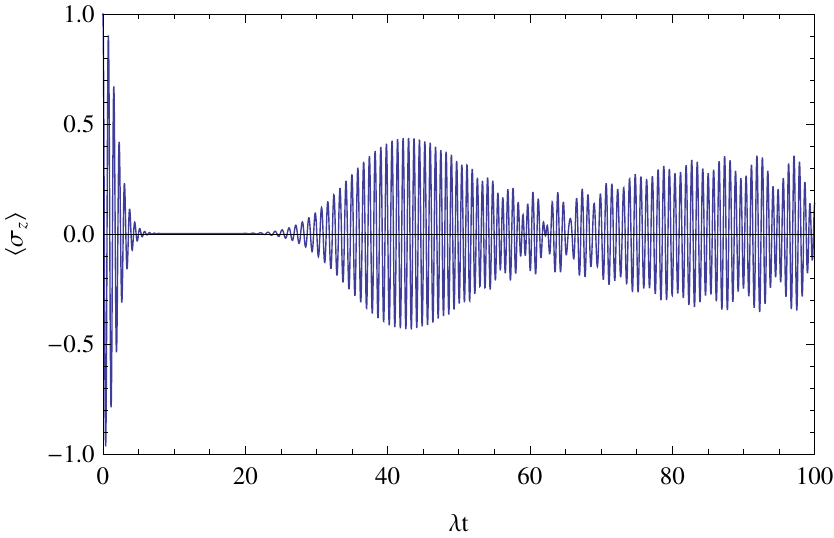}}\\
\subfloat[]{\includegraphics[scale=0.8, natwidth=640, natheight=480]{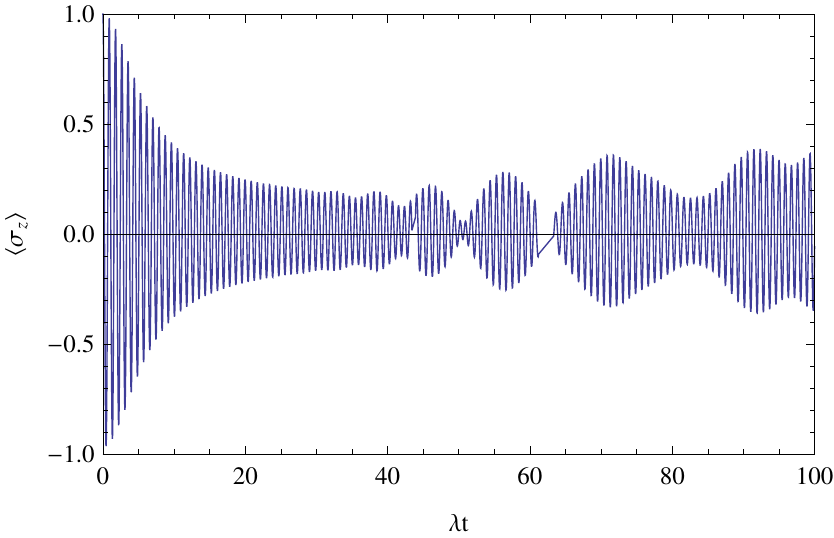}}%
\caption{Plots of the temporal evolution of the atomic inversion $\left\langle \hat{\sigma}%
_{3}\left( t\right) \right\rangle $ in the spin model in the exact resonant case. The atom is initially in the excited state and the field is initially prepared in the $SU(2)$ coherent state with  $\left\langle \hat{n}\right\rangle =20$ photons on average. Frames a), b) and c) correspond to $2j=1000, 100$ and $50$ maximum number of excitations, respectively}
\end{center}
\end{figure}

\subsection{Photon antibunching}

Both theoretically and experimentally, there is an interest in a variety of
statistical properties of the electromagnetic field, including the
distributions of possible field energies, the photon number variance and the
Mandel $\mathcal{Q}$ parameter (normalized second factorial moment) \cite{28}%
:%
\begin{equation}
\mathcal{Q}\left( t\right) =\frac{\left\langle \left( \Delta \hat{n}\right)
^{2}\right\rangle -\left\langle \hat{n}\right\rangle }{\left\langle \hat{n}%
\right\rangle }.  \label{MANDEL}
\end{equation}%
These statistical variables offer quantitative measures of how much the
field differs from a classical field. In particular, the Mandel $\mathcal{Q}$
parameter vanishes for a Poissonian distribution. It provides information
about the tendency of photons to arrive in bunches: when $\mathcal{Q}>0$ the
photons are bunched (super-Poisson), while for $\mathcal{Q}<0$ the photons
are antibunched (sub-Poisson) \cite{27}. It is well know that a\textbf{\ }%
sub-Poissonian statistics is a signature of the quantum nature of the field.

Averages appearing in Eq.(\ref{MANDEL}) can be obtained easily from Eq.(\ref%
{MEANVAL}). From figure 3(a) we can see that the photon number distribution
oscillates between sub-Poissonian and super-Poissonian statistics when $%
2j>>1 $. Figures 3(b) and (c) show that as $2j$ decreases, $\mathcal{Q}$
becomes negative and the photons are antibunched. From these results we
infer the purely quantum regime of the spin model for the electromagnetic
field.

\begin{figure}%
\vspace {0.5 cm}
\begin{center}
\subfloat[]{\includegraphics[scale=0.8, natwidth=640, natheight=480]{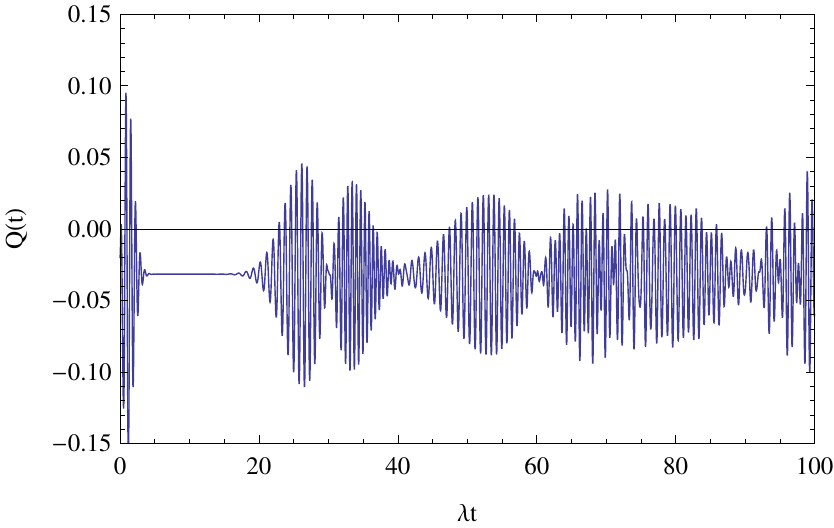}}\qquad
\subfloat[]{\includegraphics[scale=0.8, natwidth=640, natheight=480]{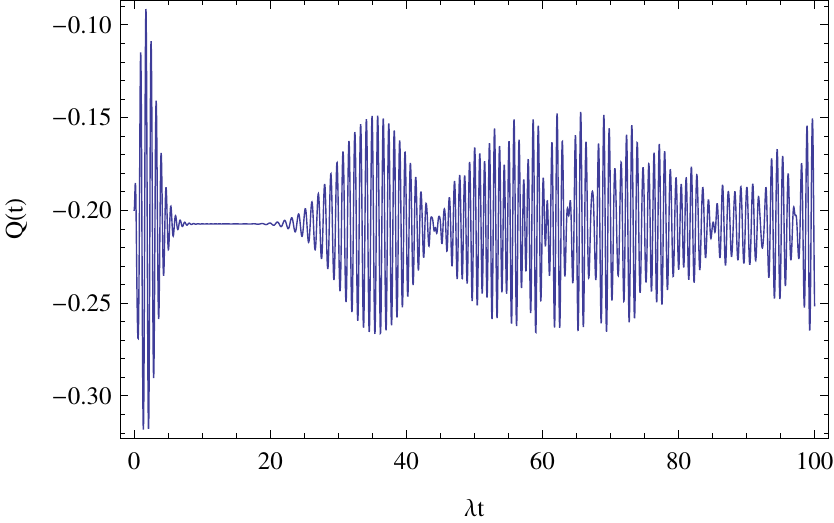}}\\
\subfloat[]{\includegraphics[scale=0.8, natwidth=640, natheight=480]{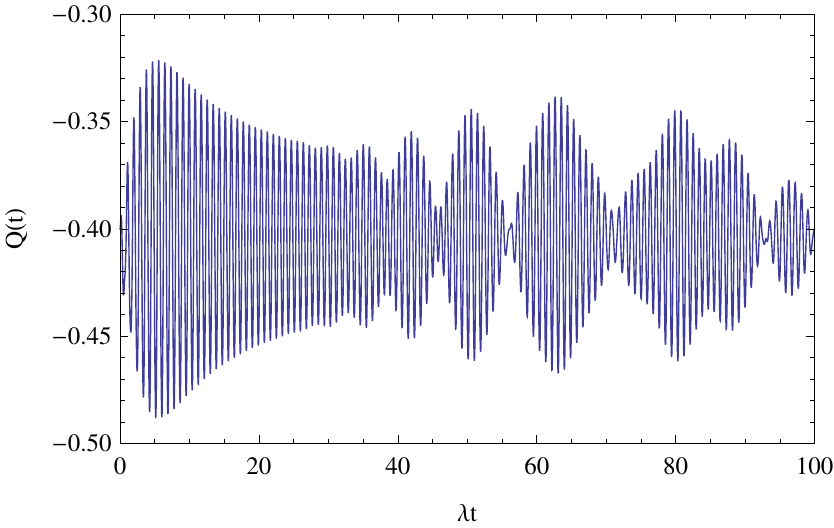}}%
\caption{Plots of the temporal evolution of the Mandel $\mathcal{Q}\left( t\right)$ parameter in the spin model in the exact resonant case. The atom is initially in the excited state and the field is initially prepared in the $SU(2)$ coherent state with  $\left\langle \hat{n}\right\rangle =20$ photons on average. Frames a), b) and c) correspond to $2j=1000, 100$ and $50$ maximum number of excitations, respectively}
\end{center}
\end{figure}

\subsection{Squeezing}

To analyze the squeezing properties of the radiation field we introduce two
hermitian quadrature operators:%
\begin{equation}
\hat{x}=\frac{\hat{J}_{x}}{\sqrt{2j}}\text{, \ \ \ \ \ \ \ \ \ \ \ }\hat{y}=%
\frac{\hat{J}_{y}}{\sqrt{2j}}\text{.}  \label{QUAD}
\end{equation}%
One of the consequences of the commutation relation for these operators is
the uncertainty relation $\left\langle \left( \Delta \hat{x}\right)
^{2}\right\rangle \left\langle \left( \Delta \hat{y}\right)
^{2}\right\rangle \geq \left\vert \frac{{\hbar }^{2}}{4}\left( 1-\frac{%
\left\langle \hat{n}\right\rangle }{j}\right) \right\vert ^{2}$. When either
of these variances is less than $\frac{{\hbar }^{2}}{4}$ (uncertainty
associated with the coherent field, including the vacuum), the state is said
to be squeezed \cite{29, 30}. In the usual JCM, the cavity field surrounding
a two-level atom initially excited, exhibits a time-varying pattern of
squeezing both in the short time regime and during the revivals \cite{4A, 32}%
. Thus the dynamics of the JCM leads to the squeezing of the radiation
field, although the effect is rather weak.

The variances of the quadrature operators can be expressed through the mean
values of the spin operators%
\begin{eqnarray}
\left\langle \left( \Delta \hat{x}\right) ^{2}\right\rangle &=&\frac{1}{2j}%
\left[ \frac{1}{2}\mathrm{{Re}\left\langle \hat{J}_{+}\hat{J}%
_{-}\right\rangle +\frac{1}{2}{Re}\left\langle \hat{J}_{+}^{2}\right\rangle
-\left( {Re}\left\langle \hat{J}_{+}\right\rangle \right) ^{2}}\right] \text{%
,} \\
\left\langle \left( \Delta \hat{y}\right) ^{2}\right\rangle &=&\frac{1}{2j}%
\left[ \frac{1}{2}\mathrm{{Re}\left\langle \hat{J}_{+}\hat{J}%
_{-}\right\rangle -\frac{1}{2}{Re}\left\langle \hat{J}_{+}^{2}\right\rangle
-\left( {Im}\left\langle \hat{J}_{+}\right\rangle \right) ^{2}\,\,}\right] 
\text{.}  \nonumber
\end{eqnarray}

In figure 4 we plot the temporal evolution of the uncertainties as a function of the scaled time. We observe in figure 4(a) that $\left\langle \left( \Delta \hat{x}\right) ^{2}\right\rangle $ exhibits squeezing periodically in both cases, for $2j=1000$ (blue line) and $50$ (red line) maximum number of excitations. In figure 4(b) it can be seen that $\left\langle \left( \Delta \hat{y}\right)^{2}\right\rangle $ is larger than the minimum even at $t=0$. On the other hand, figures 4(c) and (d) show that in the longer time behavior the uncertainties remain bounded in both cases (for $2j=1000$ and $50$) . However, they always remain significantly larger than the minimum value and thus the field is no longer squeezed.

\begin{figure}%
\vspace {0.5 cm}
\begin{center}
\subfloat[]{\includegraphics[scale=0.8, natwidth=640, natheight=480]{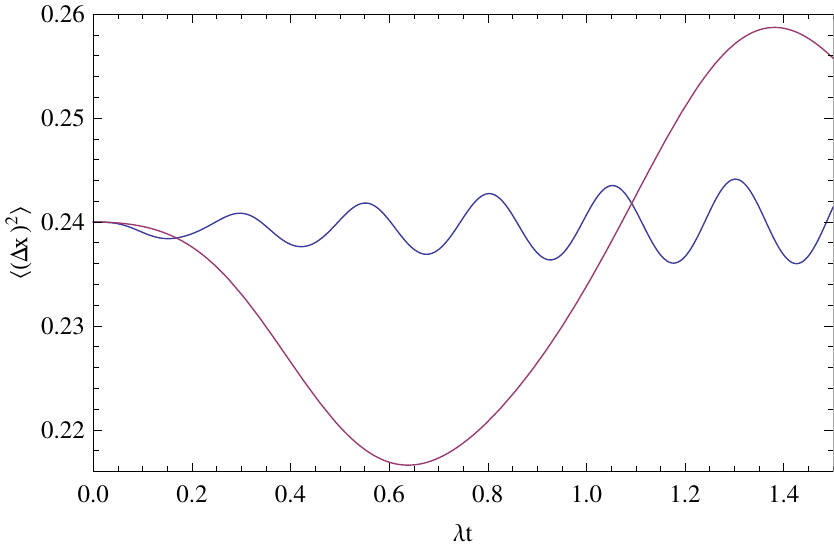}}\qquad
\subfloat[]{\includegraphics[scale=0.8, natwidth=640, natheight=480]{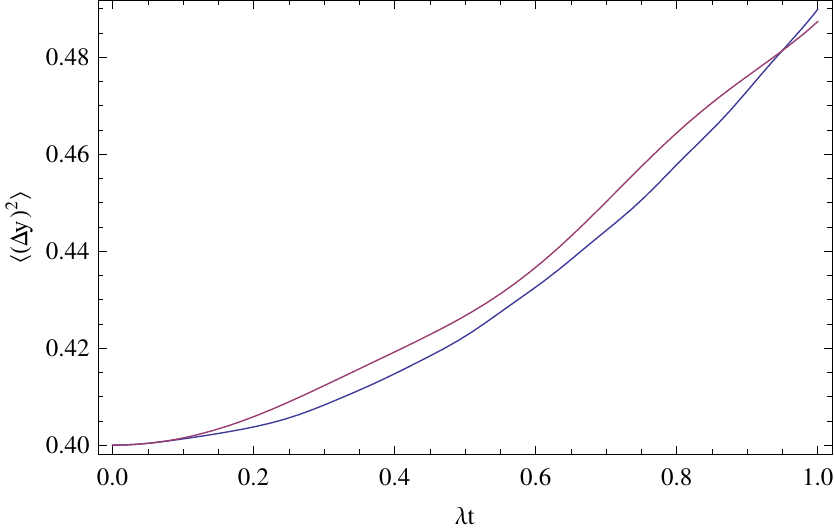}}\\
\subfloat[]{\includegraphics[scale=0.8, natwidth=640, natheight=480]{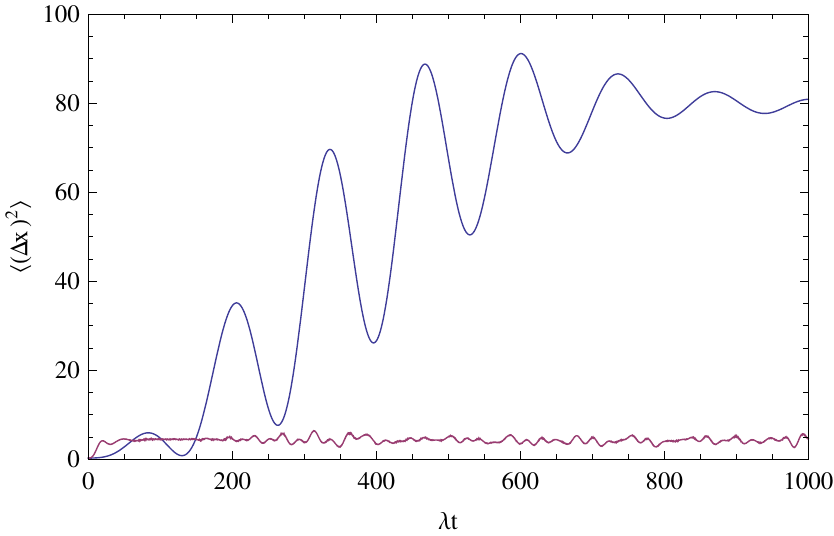}}\qquad
\subfloat[]{\includegraphics[scale=0.8, natwidth=640, natheight=480]{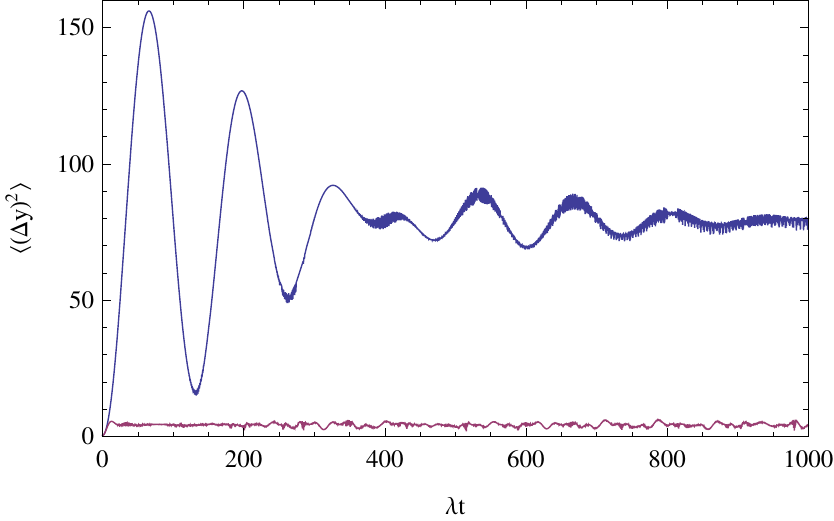}}%
\caption{Temporal evolution of the uncertainties as a function of the scaled time. Frames a) and b) show the short time regime of $\left\langle \left(
\Delta \hat{x}\right) ^{2}\right\rangle $ and $\left\langle \left( \Delta \hat{y}\right) ^{2}\right\rangle $, respectively. In the same order, frames c) and d) show the longer time behavior. The atom is initially in the excited state and the field is initially prepared in the $SU(2)$ coherent state with  $\left\langle \hat{n}\right\rangle =20$ photons on average. The maximum number of excitations are $2j=1000$ (blue line) and $2j=50$ (red line).}
\end{center}
\end{figure}

\section{Conclusions}

In this work we have introduced a spin model which exhibits the main
properties of a Kerr medium to describe an intensity dependent coupling
between a two-level atom and the radiation field. The model is formulated in
terms of spin operators acting on a finite dimensional Hilbert space, so
that the number of excitations of the field is bounded from above. We have
analyzed the behavior of both the atomic and the field quantum properties
when the atom is initially in the excited state and the field is initially
prepared in the $SU\left( 2\right) $ coherent state in the exact resonant
case.

It has been shown that the atomic population inversion exhibits periodic
collapses and revivals when $2j\gg 1$, while with decreasing $2j$ the
structures start to deteriorate. When the mean photon number is maximal ($%
\chi =1$), the atom remains in the excited state. As regards to the quantum
properties of the field, we showed that the photon number distribution
oscillates between sub-Poissonian and super-Poissonian statistics when $2j>>1
$, while as $2j$ decreases $\mathcal{Q}$ becomes negative and the photons
are antibunched. In the same fashion, we find squeezing only in the short
time regime. We are currently exploring applications of a field-theoretical
extension of these ideas \cite{33}.

\section*{Acknowledgements}

L.F.U is partially supported by project DGAPA-UNAM-IN111210. He also
acknowledges the hospitality at Facultad de F\'{\i}sica, PUC. Alejandro
Frank acknowledges support from the projects DGAPA-UNAM-IN114411 and
CONACYT-155663.

\end{document}